# Searching with Quantum Computers

*Lov K. Grover*
***Bell Labs, 700 Mountain Avenue, Murray Hill NJ 07974***

*The processing speed of quantum computers is predicted to be far superior to their classical counterparts for a few important applications. This paper introduces quantum computers by using the quantum search algorithm as an example.*

**How small can transistors get?** When solid-state transistors were first invented in 1947, they were a few cm. wide. Since then, the sizes of transistors have been shrinking steadily. In 1976, Gordon Moore of Intel quantified this trend by proposing that the size of transistors would keep shrinking at such a rate that the area of transistors would keep falling by a factor of 2 every eighteen months. This came to be known as Moore's law and has closely predicted the evolution of transistor sizes in the last two decades.

The transistors, although based on quantum mechanics, are classical information processing devices - each node is either a 0 or a 1. As devices get smaller, quantum effects become more and more significant. In fact, in present day transistors, quantum mechanical effects are already important and must be considered so as not to let them affect the working of devices. Soon, it will be necessary to harness these effects. The question arises as to whether it will still be possible to design working circuits. The answer is that it will still be possible; however, the principles of design in the new regime will be very different. One way would be to redesign the circuits so that the information itself is processed quantum mechanically. Such a device would be a quantum computer. It has been shown that anything that can be computed by a classical computer, could be computed equally efficiently with a quantum computer with a comparable amount of hardware. This is significant since it shows that quantum effects do not pose an insurmountable barrier to computer design.

Quantum mechanical computers were first proposed in the early 1970's. In these, each node, instead of being in a 0 state or a 1 state, can be in both states simultaneously. Therefore it is possible to carry out multiple computations at the same time and with proper design it is possible to take advantage of this parallelism. Recently two such quantum algorithms for the tasks of factorization and search have been invented. This has generated a lot of interest in quantum computers among both physicists and computer scientists.

Quantum hardware poses challenges of its own, not only because it must deal with microscopic particles that behave differently from classical objects, but also because quantum mechanical systems are very delicate; the slightest external perturbation affects the computation. The hardware must be extremely well isolated from the environment. So far, researchers have only been able to build machines with about 7 quantum bits operating at once, and even that only for a few microseconds.

**From probabilistic to quantum systems** Quantum mechanics is a theory about the properties of very small things. For those with an engineering background, this theory is most easily grasped as an extension of classical probability.

In order to describe a classical probabilistic system, we need to specify the probabilities in each state. Quantum mechanical systems have a deeper structure than classical systems. In addition to having a certain probability of being in each state, quantum mechanical systems also have a *phase* associated with the probability in each state which leads to wavelike interference as the system evolves. The complete specification of the state of the system requires the specification of both the magnitude as well as the phase of this probability. This is usually specified as a complex number called the amplitude - the absolute square of this complex number gives the probability of the system being in the respective state. The complete description of a quantum mechanical system is given by its *amplitude vector*, i.e. the specification of the amplitude in each state (the amplitude vector is also known as the *wavefunction* or the *superposition*).

The dynamics of a quantum mechanical system is based on the amplitude vector. Just as the evolution of a probabilistic system is described by considering the initial probabilities and the transition probabilities between various states, the evolution of a quantum system is obtained by considering the initial amplitudes and the transitions between various states.

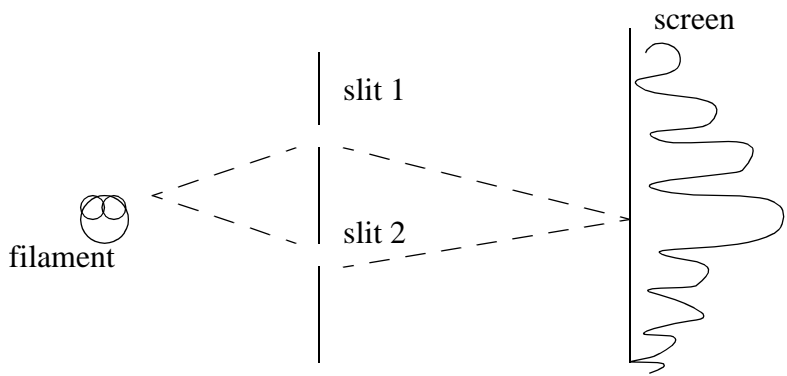

**Figure 1 - In the two slit experiment, electrons are sent in through the two slits and the distribution on the screen is observed.**

**The initial state of the quantum mechanical system is specified by the amplitudes of an electron at the two slits. The evolution of the system (e.g. the distribution of the electron at the screen) is determined by these amplitudes.**

**Bits & Qubits** Just as classical Boolean circuits are synthesized out of two-state systems called bits, quantum mechanical circuits are synthesized out of two-state quantum mechanical systems called qubits. The difference is that a bit is in either a 0 or a 1 state, whereas a qubit is in both 0 *and* 1 states at the same time with different amplitudes in the two states. To completely specify a classical bit requires us to specify just one bit of information. To specify a qubit requires us to specify the amplitudes in both states as two complex numbers. For example: (0.707 exp($i\pi/4$), 0.707 exp($-i\pi/4$)) is an amplitude vector that specifies the amplitudes in the 0 and 1 states respectively. The probabilities in the two states are the absolute squares of these amplitudes, i.e. the probabilities in each of the 0 and the 1 state is 0.5. Since the sum of the probabilities in the two states is unity, it indicates that we have a valid amplitude vector.

The analysis of quantum systems is based on the fact that the evolution of quantum systems is linear in the amplitudes. What this means is, that if the amplitude vector $(\alpha_1, \beta_1)$ evolves into $(\alpha_2, \beta_2)$, and under the same physical conditions the amplitude vector $(\gamma_1, \Delta_1)$ evolves into $(\gamma_2, \Delta_2)$, then the amplitude vector: $(\alpha_1 + \gamma_1, \beta_1 + \Delta_1)$ must evolve into $(\alpha_2 + \gamma_2, \beta_2 + \Delta_2)$. The fact that this linearity must hold for all valid values of αs and βs, along with reasonable physical assumptions, leads to most of the structure of quantum mechanics.

One aspect of this structure follows from the fact that any transition must conserve the overall probability. This, along with the linearity requirement leads to the constraint that only the class of *reversible* boolean gates can be synthesized through quantum computing. A reversible gate is one for which it is possible to deduce what the values of the inputs were from the outputs. An example of a reversible gate is NOT; an example of a gate that is not reversible is NAND.

Fortunately, any Boolean function that can be computed by standard NAND and NOR gates, can also be computed by a circuit which is of roughly the same size and composed entirely of reversible gates such as NOT and XOR. Thus, as mentioned in the introduction, any Boolean function that can be computed classically, can be equally easily computed quantum mechanically.

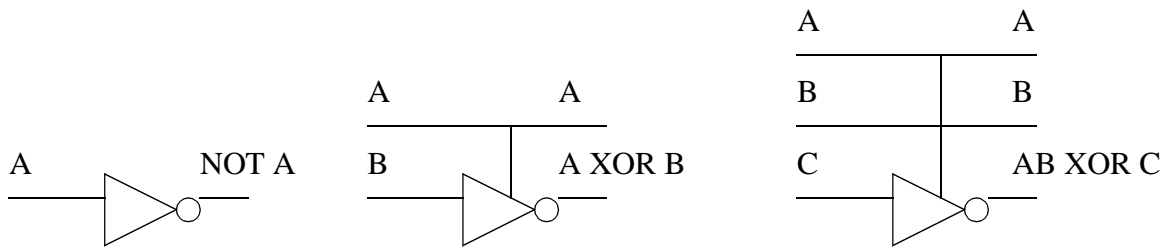

**Figure 2 - A reversible set of primitives in terms of which any Boolean function f(x) can be computed with roughly the same number of gates as with classical NANDs and NORs.**

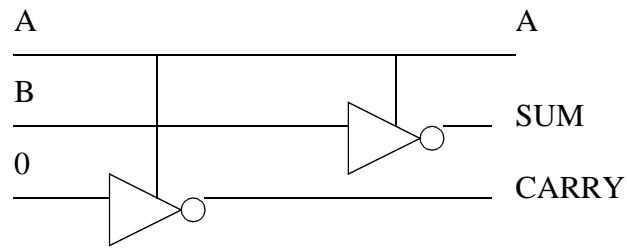

**Figure 3 - An adder circuit that uses reversible components of the type shown in the previous figure.**

**Measurement & Collapse** The characteristic of multi-particle quantum systems that leads to many of their puzzling effects is the *collapse* of the wavefunction. Whenever any component of the wavefunction is observed, the rest of the wavefunction immediately readjusts itself so as to be consistent with the observation.

Some of the consequences of this were so puzzling that Einstein at one time thought he had disproved quantum mechanics. However, after some thought, Bohr pointed out that the results, though paradoxical, did not violate any physical law. Since then quantum mechanics has survived more than six decades of intense threoretical and experimental scrutiny.

As far as quantum computation is concerned, this means that it is not allowed to externally observe a system in the middle of its computation. The algorithm has to be designed in such a way that it carries out its computation completely, only after which the observation is made. In case intermediate observations are made, their effect on the computation must be carefully analyzed (intermediate observations are not used in the quantum search algorithm).

**Algorithms** In the mid 1930's, a British mathematician, Alan Turing, showed that if something could be computed on a classical computer it could be computed on a different classical computer with approximately the same amount of resources. Now, the complexity of different problems could be studied without regard to what computer they would be implemented on. This gave rise to a theory of computation. There has been tremendous work in this area in the last half-century; however the foundations of this theory have stayed virtually unaltered.

In the early 1980's, the famous physicist Richard Feynman, in a pioneering paper, observed that there were certain problems that could easily solved by any quantum computer, yet were very difficult for a classical computer. During the mid 80's and early 90's several such problems were studied. This showed that all the previously known results of the theory of computation were merely approximations to the *real* theory of computation just as Newton's theory of gravitation had been an approximation to the real theory which was only discovered by Einstein several centuries later.

As mentioned previously, one of the constraints with quantum mechanical operations is that each operation needs to be reversible. This means that in quantum algorithms too this needs to be kept in mind and each statement needs to be reversible. Common statements like "a = 5" can be used during initialization but are not allowed within the quantum program since they irreversibly overwrite the value of a. On the other hand, statements such as "a = a + 5" are allowed since if we are given the final value of a, it is possible to deduce what the initial value was.

The potential advantage of quantum computing is the inbuilt parallelism which allows multiple computations to be performed simultaneously. The problem is that even though after this simultaneous computation, all the answers are in the system, there is no easy way of extracting all this information. For a long time it was not known whether quantum computers could solve any problem of practical interest faster than classical computers. After all, if classical computers could solve any problem of interest as fast as quantum computers, it was really not worth the effort of designing and building quantum computers. That has changed in the last five years with the discovery of fast quantum computing algorithms for factorization and searching.

**The Quantum Search Algorithm** In the exhaustive search problem, a function $f(x)$, $x = 0, 1, \ldots (N-1)$, is given. The only thing we know about $f(x)$ is that it is a binary function that is $0$ everywhere except at a single value of $x$, where it is $1$. The goal is to find the value of $x$ for which $f(x) = 1$. Clearly, any classical algorithm will need $O(N)$ steps. However, as mentioned earlier, quantum computers can be in multiple states and carry out multiple computations at the same time - if we were given a quantum mechanical circuit that evaluated $f(x)$, there is no clear bound as to how fast the search could be done. It was shown in 1994, by using subtle properties of quantum mechanical operations, that even a quantum computer would need at least $O(\sqrt{N})$ steps to carry out the search. Two years later in 1996, I discovered the quantum search algorithm that took precisely $O(\sqrt{N})$ steps.

The quantum search algorithm starts out by setting the system to a superposition of $N$ states corresponding to the $N$ items to be searched. It can now examine all $N$ items in a single step. However, if it is programmed to immediately print out the item examined, it will print a random item and therefore it will only print out the right item with a probability of $\frac{1}{N}$. Instead, by carrying out a set of quantum mechanical operations, it is possible to increase the amplitude and hence the probability in the desired state at the expense of other states. After this, it will indeed print out the right value of $x$ with a high probability.

The steps required to increase the amplitude in the desired state are easiest understood by first thinking of the following simple classical probabilistic algorithm which works similarly, by increasing the probability with each iteration. After this, we will talk about how to increase probability amplitudes with each iteration in a quantum computer; the general idea is similar.

```
int random();              /* random(N) returns random number between 0 and (N-1) */
int f();                   /* f(r) is an arbitrary binary function that returns 1 only
                              for a single value of r, for all other values of r it returns 0 */
main()
{
    int i, r, answer = -1;
    r = random(N);
    for(i = 0; i < N; i++)
    {
        if (f(r) == 1)   answer = r;
        r = random(N);
    }
    print(answer);
}
```

The success probability increases by approximately $\frac{1}{N}$ in each iteration of the for loop and therefore in $O(N)$ iterations, it will find the solution with a high probability.

The algorithm as it stands above will not work quantum mechanically. The reason is that each component of a quantum algorithm needs to be reversible. In the above algorithm, the statement: "if (f(r) == 1)   answer = r" is not reversible since if we are given the value of all the variables after the statement, it may not be possible to determine what the variables were before the statement. Also, the statement: "r = random(N)" is not reversible because once the value of r gets overwritten, it is not possible to determine what the original value of r was. As mentioned in an earlier section, it is always possible to design a quantum mechanical algorithm that accomplishes approximately the same as any classical algorithm with roughly the same hardware. The quantum algorithm will work in terms of amplitudes which are analogous to probabilities but can be either positive or negative.

It is possible to have a quantum random number generator, but in order for this to be reversible, it will need to depend on the initial state of the system. This sounds contradictory because a random number generator is supposed to generate all numbers with equal probability, irrespective of the initial state of the system. The way out of this contradiction is to let the quantum random number generator return all numbers with equal probabilities, but with the amplitudes having different signs. The information about the initial state is stored in the signs. If these signs are prop-erly chosen, the initial state can be recovered and the random number generator can indeed be made reversible even though each number is returned with the same probability. The following paragraph describes one such choice of signs (in quantum computing jargon, this random number generator is known as the Walsh-Hadamard Transformation although in this article we denote it by qrandom() ).

qrandom(N,r) generates a number, say q, between 0 and (N-1) with an amplitude whose magnitude is $1/\sqrt{N}$ for each number. The sign of this amplitude is positive or negative and determined by the following calculation. Consider the binary representation for q and r. If the number of positions at which the binary forms of *both* q and r are 1 is even, the sign of the amplitude is positive; if odd, the sign is negative. For example, if q=01110101 and r=10110111, then the places where q and r, both have 1's are (counting from the least significant bit): 1st, 3rd, 5th and 6th. Therefore the number of 1's at common positions is 4 which is even, therefore the sign of the amplitude will be positive. It can be shown that the operation "r = qrandom(N,r)" can be reversed by repeating it.

In addition to the quantum random number generator, the other quantum mechanical operation we use is a phase inversion operation. This keeps the probability in each state the same but inverts the phase of the amplitude in certain states. It is easily seen that a phase inversion operation can be reversed by applying it twice.

Using these operations, it is possible to solve the search problem of size $N$ in only $O(\sqrt{N})$ steps. The following quantum algorithm accomplishes this. It may be verified that the algorithm is composed entirely of reversible operations.

```
int qrandom();              /* qrandom(N,r) returns a random number between 0 and (N-1) */
int f();                    /* f(r) is an arbitrary binary function that returns 1 only
                               for a single value of r, for all other values of r it returns 0 */
quantum_main()
{
    int i, r;
    r = qrandom(N,0);
    for(i = 0; i < eta; i++)              /* eta is a number O(√N) */
    {
        if (f(r) == 1)    invert_phase();
                                          /* f(r) is evaluated within the program itself
                                             this does not need an external observation
                                             (any observation would perturb the system) */
        r = qrandom(N,r);
        if (r == 0)       invert_phase();
        r = qrandom(N,r);
    }
    print(r);
}
```

It must be emphasized that the above algorithm is inherently quantum mechanical and, as it stands, *cannot* be implemented on classical hardware. The fundamental difference between this and any classical probabilistic algorithm, is that this works in terms of amplitudes which can be either positive or negative, whereas the classical algorithm works in terms of probabilities which are required to be positive.

Why does the quantum mechanical algorithm require only $O(\sqrt{N})$ iterations? The analysis of the algorithm is harder than that of the classical algorithm and based on the properties of the function qrandom(). Rather than give a general analysis, the next section considers a special case. The rough idea as to why the quantum algorithm needs only $O(\sqrt{N})$ iterations is presented in the next paragraph.

When the classical system is initialized by a randomization operation, there is a $\frac{1}{N}$ probability for each value - as a result each iteration of the classical algorithm increases the probability of the right answer by approximately $\frac{1}{N}$. Similarly when the quantum system is initialized, the probability for each value of r is $\frac{1}{N}$; therefore the amplitude for each $r$, which is the square root of the probability, is $\frac{1}{\sqrt{N}}$. Since the quantum system works with amplitudes, each iteration of the "for" loop can increase the amplitude of the right answer by approximately $\frac{1}{\sqrt{N}}$. The phase inversion operations ensure that the amplitudes for transitions into the right answer add up while the amplitudes for transitions into other possibilities cancel out. Indeed in $O(\sqrt{N})$ iterations of the "for" loop, the amplitude of the right answer rises to a number of order 1. The probability being the square of the amplitude also rises to order 1.

**The N=4 case:** In general, eta (the number of iterations required by the quantum search algorithm), is a number between $0.5\sqrt{N}$ and $0.8\sqrt{N}$. Its precise value can be analytically obtained. The precise value is important for the correct working of the algorithm because, unlike in the classical case, the success probability is not monotonic with the number of iterations. The success probability oscillates between 0 and 1 with a certain periodicity, if we stop exactly at the right number of iterations the success probability is close to 1.

For the N=4 case, eta is exactly 1 and the "for" loop in the algorithm needs to be executed just once. The steps of this program are:

```
r = qrandom(4,0);
if (f(r) == 1) invert_phase();
r = qrandom(4,r);
if (r == 0)    invert_phase();
r = qrandom(4,r);
print(r);
```

As mentioned in the previous section, qrandom(N,r) generates a random number between 0 and (N-1) with an amplitude of $\pm\frac{1}{\sqrt{N}}$. In particular, qrandom(4,r) returns a number (q) in the range [0,3] with amplitude $\pm 0.5$. The sign depends on r and q as shown in figure 4:

|   | r → |   |   |   |
|---|---|---|---|---|
| q ↓ | 0 | 1 | 2 | 3 |
| 0 | 0.5 | 0.5 | 0.5 | 0.5 |
| 1 | 0.5 | (-0.5) | 0.5 | -0.5 |
| 2 | 0.5 | 0.5 | -0.5 | -0.5 |
| 3 | 0.5 | -0.5 | -0.5 | 0.5 |

**Figure 4 - The quantum mechanical random number generator qrandom(4,r) generates a number (q) between [0-3] with an amplitude either +0.5 or -0.5. This figure shows these amplitudes for various values of q and r.**

In figure 4, the sign of the amplitudes is calculated as described in the previous section by counting the number of positions at which the binary forms of *both* q and r are 1. For example, consider the case when q and r are both 1 (encircled in the above figure). The binary form of q (01) and the binary form of r (01) both have a 1 in the 2nd position. Since the number of 1's in the same positions is odd, the sign of the amplitude is negative and the amplitude of qrandom(4,1) returning a 1 is -0.5.

    With the information in figure 4, it is a straightforward, though a tedious, exercise to analyze quantum_main(). The analysis as shown in figure 5, consists of calculating the amplitude vectors for the system (in this case r) at each step of the program. This is just as when debugging a classical program we would list out the value of all variables at each step. The difference is that a classical variable assumes only one value and we just need to list that one value. On the other hand, a quantum system is in all states with certain amplitudes and we need to list out the amplitudes for all possible states.

| **Steps of the program quantum_main()** | **Amplitude vector for r** |
|---|---|
| r=qrandom(4,0);  (i) |  |
|  | (0.5, 0.5, 0.5, 0.5) |
| if (f(r) == 1)  invert_phase();  (ii) |  |
|  | (0.5, 0.5, -0.5, 0.5) |
| r = qrandom(4,r);  (iii) |  |
|  | (0.5,- 0.5, 0.5, 0.5) |
| if (r == 0)  invert_phase();  (iv) |  |
|  | (-0.5, -0.5, 0.5, 0.5) |
| r = qrandom(4,r);  (v) |  |
|  | (0.0, 0.0, -1.0, 0.0) |

**Figure 5 - The evolution of amplitudes of r during the steps of the quantum search algorithm when N=4 and the solution that makes f(x)=1 is the third value.**

As shown in figure 5, after carrying out the steps of the quantum_main() program, the amplitude of the correct number is -1 (the probability is hence 1) and the amplitudes of all other numbers is 0 (the probabilities are hence 0). An observation now reveals the right answer with certainty. A single evaluation of f(x) (in step (ii)), along with some preprocessing and postprocessing operations, suffices to identify the answer with certainty in the N=4 case. In the classical case, it would take 3 evaluations of f(x) to locate the correct number. A similar analysis for the general case shows that it can identify the answer in only $O(\sqrt{N})$ steps. Also, the quantum algorithm is able to cancel the amplitudes in all the incorrect possibilities since certain transition amplitudes are negative - hence it can find the correct answer with certainty. A classical probabilistic algorithm, like the one described in the previous section, will have a finite error probability.

The following is an outline of the calculations of the amplitude vector for r at each step of the program:

(i) The first step is qrandom(4,0). According to figure 4, this creates an amplitude of +0.5 for each of the four values of r. The amplitude vector becomes (0.5, 0.5, 0.5, 0.5)

(ii) Next, the amplitude is inverted for the value of r satisfying f(r)=1. Assuming this is the third number, the amplitude vector becomes (0.5, 0.5, -0.5, 0.5). This is the only step that requires an evaluation of f(x).

(iii) After this, the operation r = qrandom(4,r) is carried out which seems to randomize r; however, as mentioned earlier, the transition amplitudes have carefully chosen signs that preserves the information. After (iii), r can assume any value in the range [0-3] with certain amplitudes that need to be calculated in a way very similar to those for probabilities. To illustrate this, the amplitude for r to be 0 after step (iii) is calculated:

There are four paths through which r can reach 0 after (iii). To estimate this amplitude, we need to sum the amplitudes over these four paths. The four paths are:
(a) r was 0 before step (iii) and stays unchanged during step (iii).
(b) r was 1 before step (iii) and step (iii) causes it to make a transition from 1 to 0.
(c) r was 2 before step (iii) and step (iii) causes it to make a transition from 2 to 0.
(d) r was 3 before step (iii) and step (iii) causes it to make a transition from 3 to 0.

Just as for probabilities, we calculate the amplitudes of each of these four paths and sum them to find the total amplitude. The amplitude of (a) is given by multiplying the amplitude of r being initially 0 (0.5) by the amplitude of it staying 0 during step (iii) (this is read off from figure 4 and is 0.5). Thus the amplitude of (a) is 0.5 X 0.5 which is 0.25. Similarly the amplitude of (b) is 0.25, that of (c) is -0.25, and that of (d) is 0.25. Summing these four amplitudes leads to a total amplitude of 0.5 for r to be 0 after step (iii).

Carrying out these calculations for the other 3 possible values of r leads to the result that the amplitude vector for r after step (iii) is (0.5, -0.5, 0.5, 0.5).

(iv) Next the amplitude in the 0 state is inverted and the amplitude vector for r becomes:
(-0.5, -0.5, 0.5, 0.5).

(v) Finally the operation "r=qrandom(4,r)" is applied again and as in (iii), the amplitudes of r having different values can be calculated by considering the paths through which r can reach each value. After step (v), the amplitude vector for r can be shown to be: (0.0, 0.0, -1.0, 0.0).

**Realizations** Quantum mechanical systems are very delicate and have to be designed to be almost completely isolated from their environments. This is because, as mentioned previously, whenever any portion of a quantum system is observed, the rest of the system readjusts itself so as to become consistent with the observation. Thus any time a quantum system interacts with its environment (i.e. any macroscopic system) it gets perturbed just as if someone had tried to observe it. This affects the computation in unpredictable ways.

Isolating a large system is difficult. Our environment contains a lot of background noise such as stray dust particles & stray photons which rapidly perturb any macroscopic body. For example, a football lasts less than $10^{-20}$ seconds before it is struck by a stray particle. On the other hand, tiny particles can stay isolated for much longer, e.g. an atom can stay isolated for hours or even days. The problem is that it is much harder to carry out controllable operations on a quantum mechanical system of atomic scale. Some techniques through which elementary quantum computing operations have been demonstrated are:
(i) Single photons interacting through non-linear media.
(ii) Ions in a cryogenic trap.
(iii) Nuclear magnetic resonance (NMR).
(iv) Superconducting devices in solids.

NMR is the most advanced, having demonstrated 7 qubit systems. Simple algorithms including versions of the quantum search algorithm have been implemented using NMR in organic liquids. However, considering the experience of integrated circuits, in the long term, solids appear the most promising. Unfortunately, there are several problems with solids; for example, in order to achieve isolation in solids, one needs to go to very low temperatures.

**Future prospects** Quantum computing is a field in its infancy. The two challenges in the field are the hardware & the software. The hardware challenge is to devise schemes for building quantum hardware that can naturally be isolated from the environment and on which controllable quantum operations can be carried out. The problem is that such a device will be radically different from any known device and no one knows what its structure will be like. So far only very small prototypes have been built. The software challenge is to devise applications that will make the task of building quantum hardware worthwhile. So far there are just two important applications known - search & factorization.

When will a quantum computer be built? It is interesting to recall the following quote from a visionary Popular Mechanics article of 1949 - "Where a calculator on the ENIAC is equipped with 18,000 vacuum tubes and weighs 30 tons, computers in the future may have only 1,000 vacuum tubes and weigh only 1.5 tons."

**Further reading**
(i) *Quantum computing with molecules*, Scientific American, June 1998, pages 140-145, Neil Gershenfeld and Isaac Chuang - basic introduction to quantum computing with focus on implementations.
(ii) *Quantum computing - pro & con*, John Preskill, http://xxx.lanl.gov/abs/quant-ph/9705032, an inspiring and readable description of the potential and limitations of quantum computers.
(iii) *From Schrödinger's Equation to the Quantum Search Algorithm*, Lov K. Grover, http://www.bell-labs.com/user/lkgrover/papers.html, an introductory description of the search algorithm.
(iv) *Quantum information and computation*, Charles H. Bennett & David P. DiVincenzo, Nature, Vol. 404, 16 Mar 2000, pages 247-255, a state of the art survey of quantum information.